# Research on innovation in business and management about China and Latin America: bibliometrics insights using Google Scholar, Dimensions and Microsoft Academic

Julián David Cortés-Sánchez [a,b]*; Xiaolei Lin[c]; Xiaolei Xun[c,d]

[a]*School of Management and Business, Universidad del Rosario, Bogotá, Colombia*

[b]*Fudan Development Institute, Fudan University, Shanghai, China*

[c]*School of Data Science, Fudan University, Shanghai, China*

[d]*BeiGene, Beijng, China*

*Corresponding author email: julian.cortess@urosario.edu.co

- Julián David Cortés-Sánchez is a principal professor at Universidad del Rosario's School of Management and Business (Colombia), and visiting scholar (2019) and invited researcher at Fudan University's Development Institute (China).

- Xiaolei Lin is an assistant professor in the School of Data Science at Fudan University (China). Her research interests include developing statistical methodologies for intensive longitudinal data, machine learning algorithms for complex panel data and Bayesian adaptive design in drug development. Before Fudan, Xiaolei obtained her Ph.D from the University of Chicago, under the supervision of Dr. Donald Hedeker in 2018.

- Xiaolei Xun is an Associate Director, Statistics and Data Science at BeiGene. She received a Ph.D. in Statistics from Texas A&M University, worked at Novartis for

# Research on innovation in business and management regarding China and Latin America: bibliometrics insights using Google Scholar, Dimensions, and Microsoft Academic

Trade and investment between developing regions such as China and Latin America (LATAM) are growing prominently. However, insights on crucial factors such as innovation in business and management (iBM) about both regions have not been scrutinized. This study presents the research output, impact, and structure of iBM research published about China and LATAM in a comparative framework using Google Scholar, Dimensions, and Microsoft Academic. Findings showed i) that iBM topics of both regions were framed within research and development management, and technological development topics, ii) significant differences in output and impact between regions, and iii) the same case for platforms.

## 1 Introduction

Innovation in business and management (iBM) is associated with the improvement, invention, or implementation of management practices, processes, structures, or techniques, emergent to the 'state of the art' intended to further organizational goals (Birkinshaw et al., 2008). Organizations exploring and exploiting innovation-related resources such as evidence stemming from research improve their viability in a technologically changing environment (Gao et al., 2013; Mueller et al., 2013; Pfeffer & Sutton, 2006; Rosenbusch et al., 2011; K. Zhao, 2015; S. Zhao & Yang, 2008). That factor is central for organizations in developing countries (Finardi, 2015).

Trade between developing regions such as China and Latin America (LATAM) grew by 1,200% from 2000 to 2009 (i.e., from USD$10 to USD$130 billion) (Koleski, 2011). Besides, China has poured over USD$141 billion in loans in LATAM (Jenner, 2019). Regardless of the mutual growth in trade and the large unilateral investment, research output (e.g., research articles published per year), impact (e.g., citations per paper), and structure

(e.g., cognitive proximity between research fields) on iBM about China and LATAM, have barely been studied.

A conservative number estimates that China and LATAM published 1,300+ and 1,000+ articles on iBM from 1996 to 2018, respectively (Scopus, 2018). By counting research articles by the hundreds, bibliometric methods such as citation analysis, keyword co-occurrence, and journals' co-citation networks, illuminate the view on features on research output, impact, and the underlying structure (Zupic & Čater, 2015). Studies on bibliometrics on iBM from China and LATAM presented and discussed in the 'Literature review' section entail two similarities: the dynamics between both global north and south, and the use of the bibliographic database Clarivate Analytics' Web of Science (WoS) as the pervasive bibliographic data source. Research to be published elsewhere is advancing in that front using both WoS and Elsevier's Scopus (Cortés-Sánchez, 2020). Despite the developments above, studies analyzing the relation between south-south (i.e., between developing regions or the global south) using different data sources from either WoS or Scopus have not been conducted to the best of our knowledge. Differences between databases as WoS and Dimensions vary abruptly from coverture to free access (Thelwall, 2018).

Mutual and strategic relations between China and LATAM but its lack of understanding in research on iBM and the pervasive use of only one data source to understand this dynamic, generate the following research question: what is the output, impact, and structure of the research on iBM about China and LATAM under the scope of free access bibliographic databases/search engines? Consequently, this study aims to deepen the understanding of the research output, impact, and structure of the research on iBM published between 1996-2018 regarding China and LATAM in a comparative framework using Google Scholar, Dimensions, and Microsoft Academic.

## 2 Literature review

The literature reviewed on bibliometric studies on iBM followed two pathways: geographical-comparative studies and specific findings from both regions. First, China's research-technological output on nanoscience surpassed that of the US (Kostoff, 2012), which did not translate automatically on impact. Countries like France, Germany, Japan, and the US account for most of the citations (Guan & Ma, 2007). The pharmaceutical sector's internationalization dynamics showed a defined influence from the US towards European companies' top-ten. Also, foreign affiliates and external partners of more than 50% of European companies participated in the research production (Tijssen, 2009). Comparing biotechnology clusters' labor market of the UK and Germany, the former has a balanced composition of scientists with industry and academic experience, while the latter was composed mostly of academic scientists (Casper & Murray, 2005). Harzing and Giroud (2014) proposed the comparative advantage of 34 countries' research for 21 disciplines. Five groups of countries presented a similar comparative advantage of research. For instance, groups two and five, composed of France, Italy, Poland, Russia, and Ukraine, showed an advantage in physical sciences. Studies for determining research impact made a case for factors such as previous citation impact, international collaboration, and scientific output (Confraria et al., 2017; de Paulo et al., 2017). Research agenda on technology and innovation management (TIM) processes (Cetindamar et al., 2009; Pilkington & Teichert, 2006) found that geographic regions focused on particular topics. For instance, developed countries are interested in *technology strategy, new product development,* and *design and innovation,* while developing countries in *technology policy, technological acquisition,* and *R&D management.* Also, it delineated some of the field's pillars (e.g., dynamic organizations, innovation process, knowledge management). Finally, Choi et al. (2012) identified comparative advantage on TIM by countries, for instance, the UK has consolidated differentiation in *social change,* while Spain in *intellectual property*, the Netherlands in *technology policy,* Germany in *entrepreneurship,* among others.

Second, studies focused on China stated that patents from sectors such as biotech cite more scientific papers, whereas patents from sectors such as ICT cite other patents more often (Guan & He, 2007). A compelling case for bibliometric studies was its use for technology foresight in national and regional policy design and applications regarding emerging technologies (e.g., solar cell industry) (Huang et al., 2014; N. Li et al., 2017; X. Li et al., 2015). Research in LATAM has followed different pathways, such as *Schumpeterian* innovation and cooperative relationships (Lazzarotti et al., 2011; Lopes & De Carvalho, 2012), innovativeness measures (De Carvalho et al., 2017), industry relations (Manjarrez et al., 2016), business models (Ceretta et al., 2016), financing on innovation (Padilla-Ospina et al., 2018), social innovation (Silveira & Zilber, 2017), and supply chain management (Tanco et al., 2018). That line of research outlined worldwide maturity on industrial innovation actors –not so in LATAM except for Brazil– but also a minor centrality of the region's supply chain management research for both output and impact. Emerging topics on financing innovation, such as financial constraints, funding sources, capital structure, venture capital, and financing of technology companies were spotted. This study aims to extend those findings related to iBM by comparing results using three free access platforms, namely: Google Scholar, Microsoft Academic, and Dimensions in an integrative framework assessing output, impact, and structure factors of the research about China and LATAM.

# 3 Methodology

## *3.1 Data*

This study gathered bibliographic data from Google Scholar (GS), Microsoft Academic (MA), and Dimensions. GS uses academic records from the Google search engine and also commercial, non-profit, institutional, or individual bibliographic databases. Gusenbauer (2019) calculated the number of documents indexed by Google Scholar in 398 million. MA data comes from web pages indexed by Bing. In 2017, it was estimated that 168 million documents were indexed with a monthly record growth of 1.3 million (Hug & Brändle, 2017). Finally, Dimensions is a new scholarly database launched by Digital Science in 2018. It includes awarded grants, patents, clinical trials, policy documents, and alternative metrics (i.e., altmetrics) information (e.g., citations on Wikipedia and in public policy documents, discussions on research blogs, or mentions on social networks), besides scholarly documents (Thelwall, 2018). Dimensions contains more than 90 million publications, 3.7 million grant objects, 34 million patents, and 9 million citations from scholarly articles (Hook et al., 2018). Its free version allows for the search of publications index and links to all the other different entities. GS and MA have restrictions such as the lack of control of the documents indexed or the resulting noisy database after data subtraction. Both are considered complementary and alternative tools for bibliometric and scientometric studies (Aguillo, 2012; Gusenbauer, 2019; Hook et al., 2018; Hug & Brändle, 2017; Martín-Martín et al., 2018; Thelwall, 2018). Application Programming Interface (API) access was applied via Publish and Perish (Harzing, 2007) for extracting bibliographic data from both GS and MA.

Table 1 summarizes the search query used in each search engine or bibliographic database and the results. Since the platforms do not allow for the search of authors' affiliation, this bibliographic information is not strictly on research published *by* researchers from China or LATAM but *regarding* China and LATAM. Therefore, the query was the following: the geographical place (China, Latin America, and the 20 LATAM countries); and innovation in English, Spanish, and Portuguese among document titles and abstracts. Results found 4,384 documents in GS, 1,337 in Dimensions, and 650 in MA.

**[Table 1]**

### *3.2 Methods*

In terms of impact and output, we implemented ANOVA (Analysis of Variance) for comparing both regions on bibliometric indicators calculated by each bibliographic database/search engine. Variables considered were: the number of authors, citations, and groups of authors and authors-citations. The authors refer to a proxy of authors' social capital (Wuchty et al., 2007). The four groups of authors were: individual, couples, trios, and packs (i.e., four or more authors). We analyze the publication ratio of China/LATAM, and the number of authors, and listed the top-ten most cited paper in each platform.

Regarding structure, we generated co-location, co-authorship, and journals' co-citations networks. Each bibliographic database/search engine provided data with a particular scope. Thus, the elaboration of co-authorship networks was available for GS, Dimensions, and MA; and co-citation networks for Dimensions. The co-location network used the whole sample of documents. A co-location network is a network of high-frequency key-terms appearing by proximity (Sinclair & Geoffrey, 2012). Co-authorship networks are one of the most tangible and reliable methods to study scientific collaboration and the creation of social networks of researchers by linking co-authors in a given document (Wolfgang Glänzel & Schubert, 2005). The scientific collaboration reflects individual mobility/research interest, economic/political dependence at the institutional or national level, degree of integration, or research activities between institutions (W. Glänzel & Schubert, 2001). A co-citation network connects academic disciplines by identifying a document's reference section and connecting the documents' research fields that appear in the references list (Small et al., 1985).

## 4 Results

Table 2 presents descriptive statistics of the variables by region provided by GS–the type of document, number of authors, citations, cites/author and cites/year–, Dimensions–citations, number of authors, RCR (Relative Citation Ratio) and FCR (Field Citation Ratio)–and MA –type, number of authors, citations, cites/author and cites/year.

There were significant differences in the number of authors, citations, and the group of authors at the $p<.05$ level in GS. Differences on number of authors for China (n=975) and LATAM (n=3,378) were significant [$F(1,4351)=667.665$, $p=.000$]. Post hoc comparisons using the Tukey HSD test indicated that the number of authors mean score for China ($\bar{x}=2.6$, $\sigma=1.5$) was different and higher than that of LATAM ($\bar{x}=2.0$, $\sigma=1.0$). Differences on citations for China (n=975) and LATAM (n=3,378) were significant [$F(1,4351)=163.230$, $p=.000$]. Post hoc comparisons using the Tukey HSD test indicated that the number of citations mean score for China ($\bar{x}=376.3$, $\sigma=835.6$) was different and higher than that of LATAM ($\bar{x}=4.7$, $\sigma=14.4$). Differences on number of authors-citations [G1: individual (n=1,595); G2: couples (n=1,186); G3: trios (n=989); G4: packs (four or more authors, n=583)] were significant also [$F(3,4349)=3.690$, $p=.011$]. Post hoc comparisons using the Tukey HSD test indicated that the mean score for the number of authors-citations of the pack group ($\bar{x}=129.4$, $\sigma=666,06$) was significantly different and higher than the individual group ($\bar{x}=64.4$, $\sigma=320.72$).

Results from Dimensions differed substantially. There were not found significant differences on number of authors [$F(1,1200)=1.489$, $p=.223$] between China (n=701) and LATAM (n=500); nor group of authors-citations [$F(1,1194)=.934$, $p=.424$] G1: individual (n=213); G2: couples (n=360); G3: trios (n=334); G4: packs (n=291)] at the $p<.05$ level. On the other hand, differences in citations between China and LATAM were significant [$F(1,1200)=232.366$, $p=.000$] at $p<.05$. Post hoc comparisons using the Tukey HSD test indicated that the citations mean score for China ($\bar{x}=33.66$, $\sigma=38.42$) was different, and higher, than LATAM ($\bar{x}=7.46$, $\sigma=20.59$).

Results from MA were similar to those of GS. Differences on number of authors [$F(1,647)=16.38$, $p=.000$] and citations [$F(1,647)=31.46$, $p=.000$] between China ($n=304$) and LATAM ($n=332$) were significant at the $p<.05$ level. Post hoc comparisons using the Tukey HSD test indicated that the number of authors mean score for LATAM ($\bar{x}=1.68$, $\sigma=1.1$) was different and lower than China ($\bar{x}=2.2$, $\sigma=1.4$), and that the number of citations mean score for China ($\bar{x}=5.3$, $\sigma=19.6$) was different and higher than that of LATAM ($\bar{x}=0.9$, $\sigma=4.08$). On the other hand, differences on groups of authors-citations [G1: individual ($n=328$); G2: couples ($n=142$); G3: trios ($n=100$); G4: packs ($n=78$)] were not significant [$F(3,644)=2.475$, $p=.060$] at the $p<.05$ level.

**[Table 2]**

Figure 1 presents the total output and China-LATAM for each bibliographic database/search engine for 2000-2018. There is a noticeably higher amount of publications *about* LATAM than China. It might be for search limitations, as China's results were limited to 1,000, whereas that limit was for each country from LATAM. There were three China-LATAM ratio peaks: one in Dimensions, and two in MA. In 2001, research about China was more than four times higher than LATAM in Dimensions. In MA, research about China was more than four and three times higher in 2005 and 2009. China-LATAM ratio reached a convergence (1 document about China for 1 document about LATAM) since 2008 for GS, and since 2014 for both Dimensions and MA. In the last five years, publications about China had reached an average of 40% of the publications about LATAM.

**[Figure 1]**

Figure 2 presents the ratio citations/output in periods of three years. The average citation/output (68.3) showed a decreasing trend. The average was pulled up by the citation/output ratio of China in GS. The highest citation/output ratio for China was in the period 1999-2001, with 618.8 in GS. While fluctuations along the whole period analyzed, it scored a ratio of 510 in the last two years (2017-2018). LATAM's higher performance was in Dimensions with a peak of 56 citation/output ratio in 1999-2001 and an average of 20.4. The exceptional case was in the period 2005-2007 when LATAM surpassed China with a 6.2 and 5.8 citation/output ratio in MA, respectively.

**[Figure 2]**

Figure 3 presents the increasing path in the number of authors in the bibliographic databases/search engines analyzed for 2001-2018. There is a noticeable increasing trend in the average number of authors per paper, from an average of 1.9 authors per paper in 2001 to 2.7 in 2018.

**[Figure 3]**

### *4.1 Highly cited publications*

Table 3 presents the top-five most cited documents in each region for each bibliographic database/search engine. The most cited document about China tracked by GS was a book: 'Business Research Methods' authored by Zikmund et al. (2010) and cited +10,000 times. Most of the leading authors were affiliated to either Chinese or European institutions, the rest to a North-American and LATAM institutions. Journal articles from MA and Dimensions are related to institutional and governmental effects on innovation (performance) and economic development, innovation systems comparisons, technology clusters, Global Value Chains, and spin-offs, and firm internationalization and among reputable journals, figured: *Academy of Management Review, Strategic Management Journal,* and *Research Policy.*

**[Table 3]**

### *4.2 Co-location, co-authorship and co-citations networks*

Figure 4 presents the text-mining results for all the documents' titles using Voyant Tools (Sinclair & Geoffrey, 2012). More than +167,000 words were processed to produce a co-location graph. The most innovation-related key-terms were: *performance* (1,024); *management* (959); *technology* (837); *knowledge* (793); *development* (735); *industry* (688); *firms* (663); *evidence* (640); and *product* (463). An aspect related to methodological appraisal was the emergence of both *study* (940) and *case* (910) (i.e., *case study*) terms in the corpus.

**[Figure 4]**

Co-authorship analysis in China based on GS (Figure 5), identified 13 groups of authors with at least one document published. The two largest groups of authors were composed of ten members each (red and green clusters), in which Liu, X. (China Agricultural University – H Index: 46) and Liu, Y. (Aalto University, Finland – H Index: 22) were the researchers with the highest number of co-authors. In LATAM (Figure 5), 11 groups of authors identified with at least one document published. The two largest groups of authors were composed of nine members each (red and green clusters), in which Alderete, M. (IIESS-CONICET-UNS, Argentina – H Index: 10) and Yoguel, G. (Universidad Nacional de General Sarmiento, Argentina – H Index: 33) were the researchers with the highest number of co-authors.

**[Figure 5]**

The co-authorship network based on Dimensions of China (Figure 6), identified 25 groups of authors with at least one document published. The largest group of authors was composed of 18 members (red cluster). The network highlighted Wright, M. (Imperial College, UK – H Index: 135) in terms of the number of links and connections to other clusters. In LATAM (Figure 6), there were identified three groups of authors with at least one document published. The two largest communities of authors were composed of 11 members (red), in which

Devaux, A. (International Potato Center) was the researcher with the highest number of co-authors. We omitted the results of China's co-authorship network and LATAM based on MA since they offered marginal remarks.

**[Figure 6]**

The journals' co-citation network of China based on Dimensions (Figure 7) comprised 541 documents with at least 20 citations, 75,947 links, and five clusters (>50 items): red (123 items), green (116), blue (113), yellow (79), and purple (57). Journals with high weight in each cluster were (SCImago, 2018): *Journal of International Business Studies* (H Index: 168; subject: business and international management); *Academy of Management Journal* (H Index: 283; subject: business and international management); *Journal of Operations Management* (H Index: 166; subject: strategy and management); *Journal of Marketing* (H Index: 218; subject: business and international management; marketing); and *Management Science* (H Index: 221; subject: strategy and management).

**[Figure 7]**

The journals' co-citation network of LATAM (Figure 8) was composed of 128 items, 6,179 links, and four clusters: red (45 items), green (39), blue (27), and yellow (15). Journals with high weight in each cluster were: *Research Policy* (H Index: 191; subject: management of technology and innovation); *Journal of Cleaner Production* (H Index: 150; subject: strategy and management); *Strategic Management Journal* (H Index: 232; subject: business and international management); and *Technological Forecasting and Social Change* (H Index: 93; subject: business and international management). There are no apparent similarities between China and LATAM, considering the most central journals. The co-citation network from China was diverse in terms of journals and clusters formed. The number of items and linked differences between both networks might be by the LATAM network was composed of only 23% of the items that composed China's network.

**[Figure 8]**

## 5 Discussion

Each of the bibliographic databases/search engines have their differences and similarities, reflected somehow in the bibliometric variables studied. A crucial similarity is the comprehensive inclusion of documents rather than an arbitrary selection of standards, which does apply for private Clarivate's Web of Science (WoS) or Elsevier's Scopus (Waltman & Larivière, 2020). That inclusive feature, however, varies. MA uses Artificial Intelligence (AI) that incorporates data browsing the Web, adding sources that seem to be academic and also data provided by publishers (Wang et al., 2020), while Dimensions incorporates data from Crossref and PubMed, and also input from publishers (Herzog et al., 2020). GS could reach the whole Web (!) (Falagas et al., 2007).

A consensus outcome was the significant difference in citation between China and LATAM, the former with a higher average. China's scientific wealth: production and impact (Prathap, 2017), is generated, in part, by the population and national public/private investment in R&D (King, 2004). While developing countries fall short of the 1.68% GDP world average in R&D investment, China followed the US with USD$400 billion injected into its research ecosystem, also making it the world's research production leader (Tollefson, 2018). In 2019, China's spending on science and research funding reached 2.5% of its total GDP, as it seeks to catch up on the technology front (Ng & Cai, 2019). Previously, this number had risen from 0.893% in 2000 to 2.129% in 2017. Net output findings do not seem to be supported by the convergence of the China/LATAM output ratio. We already mentioned Publish and Perish API restrictions. Furthermore, a Boolean search in GS (i.e., China AND innovation AND

business OR management) found 2.8+ million documents, which surpassed by far the 3,000+ documents gathered searching country-by-country for LATAM for the GS sample.

The difference in the number of authors per publication for both GS and MA between China and LATAM was significant, the former with a higher average. The *lonely wolf* is an endangered species: teams dominate modern research. Over the past 45 years, the authors' average number per paper went from 1.9 to 3.5 (Wuchty et al., 2007). The average number of authors in this study was 2.7 (2018), way below the global average. That is not a disadvantage *per se.* It could be a virtue, particularly for LATAM, since its average is significantly lower than that of China, with a caveat: the urgent strengthening of other research-related capabilities, resources, and incentives. Recent findings stated that small teams disrupt science and technology (Wu et al., 2019). In LATAM, however, research on innovation in business and management has been published in low-impact journals (journals' H Index below 50), and one in four papers in a journal with predatory features (Cortés-Sánchez, 2019). To incentivize academic-corporate collaboration is a driver for innovation in developing countries (Ryan, 2010), yet it is a strategy still in its infancy. In the field of innovation for sustainability in developing countries, it was reported that 1.6% (195) of documents had at least one corporate author (Cortés-Sánchez et al., 2020). The most productive corporate-academic collaboration countries were China (52%), and, from LATAM, figured Brazil (8%) author (Cortés-Sánchez et al., 2020).

Further, China's higher average could also be attributed to its encouragement for international and institutional research collaboration program (Quan Wei et al., 2017). The connection with developed countries from politics to research and development also contributed to its increasing collaborative behavior over time (He, 2009). Recently, various national talent programs that aim to recruit high-quality researchers and principal investigators facilitate international academic collaborations (Jia, 2018). However, collaborative behavior highly depends on scientific research, with more collaboration seen in physics, medicine, infectious disease, and brain sciences, and less in social science, computer science, and engineering.

Citation/output during 1999-2018 showed a decreasing trend, on average. Since mature documents have been in the academic-public domain more time than recent ones, they were exposed early to readers and potential citations. GS did not follow that trend. Further, the top-ten of GS is utterly different from MA and Dimensions. The most cited GS sources were books and book chapters while in MA and Dimensions were article journals, a remarkable difference between platforms' reach. Also, GS citations of books were 3.2 times more common than were Elsevier's Scopus citations in a sample of social and human sciences (Kousha et al., 2011). Even more in the field of innovation studies where handbooks concentrate vertical knowledge (Fagerberg et al., 2012). For the top-dive most cited documents about LATAM for each search engine/platform, the authors' affiliation was from non-LATAM institutions. That agrees with previous findings of high average citation articles led by authors from non-LATAM institutions (Cortés-Sánchez, 2019).

Previous text-mining analysis related to management and innovation, articulate with central topics also found here, such as (innovation) *management*, *technology*, and *development* (Cortés-Sánchez, 2019). Key-terms central here were relatively marginal, such as *firms*, *evidence*, or *performance.* The term *Industry* was of remarkable relevance for China's circular economy research for 2010-2013 (Cui & Zhang, 2018). None of the key topics on innovation systems research in LATAM, such as national/regional/sectoral innovation systems, were highly central here (Manjarrez et al., 2016). In essence, China and LATAM's research agenda on innovation is amid two research areas in TIM carried out by developing countries, namely *research and development management* and *technological development* (Cetindamar et al., 2009).

The co-authorship mapping features changed according to the source. While using the GS research engine dataset identified 13 groups of researchers from China, by using Dimensions, 25 groups were identified. In contrast, LATAM data from Dimensions identified just three groups, while 11 groups were identified using GS. China showed a slightly larger co-authorship cluster than LATAM in GS, while the difference is much more significant in Dimensions. China's co-authorship network tends to be more interactive than that of LATAM (more interactions between different co-authorship clusters, as shown by the edges between different nodes), indicating more diverse collaboration to produce innovative research. Each platform scope can partly explain the differences and sources. Another angle is that some authors could be more focused on publishing white/working papers or policy briefs than peer-reviewed literature, relevant differences for the reaching, and, therefore mapping, capacities of each platform (Thelwall, 2018). These findings are in line with the affiliation of central authors with universities and not with firms, public sector, or multilateral institutions, among others, since this configuration of co-authorship is still in its infancy, furthermore in developing countries, as previously stated (Cortés-Sánchez et al., 2020).

Co-citation networks highlighted new central journals in the landscape of innovation-related studies when contrasted with previous findings. Journals such as the *Academy of Management Journal, Management Science, Research Policy,* and *Strategic Management Journal*, are central and common ground (Di Guardo & Harrigan, 2012). They indicate that these journals are popular among innovation research and tend to be semantically related. Other journals emerged here. In China, such journals were: *Journal of International Business Studies, Journal of Operations Management,* and *Journal of Marketing.* For LATAM, they were: *Journal of Cleaner Production,* and *Technological Forecasting and Social Change.* The subject areas and categories of such journals are, in essence, three: business and international management, strategy and management business, and marketing. A closer look reveals that research on innovation in both regions has been permeating other areas, such as computer science applications (*Journal of Operations Management*), renewable energy and sustainability and the environment (*Journal of Cleaner Production),* and applied psychology (*Technological Forecasting and Social Change*). Previous research also showed that China has been increasingly focused on research spending in natural science and technology innovation (Veugelers, 2017). Innovation-related research in both regions is interdisciplinary, as expected (Baregheh et al., 2009), and is nurturing the mainstream (modern) innovation knowledge base, i.e., the economics of R&D, organizing innovation, and innovation systems (Fagerberg et al., 2012).

## 6 Conclusion

This study contributed to understanding bibliometric related features on the research on innovation in business and management regarding China and LATAM based on bibliographic data from GS, Dimensions, and MA. The main findings tell the significant net outcome differences and co-authorship dynamics between both regions, which is no surprise considering China's scientific wealth driven by financial and human stock invested in R&D, accompanied by international research talent recruitment and collaboration policies. The significant and higher number of authors per publication measure for China also support the former claim. A global dynamic of 'team-science' also serves as an explaining factor. GS citation counting is significantly higher, contrasting the assumption that mature documents have been in the academic-public domain more time than recent ones, which exposed them to potential citations. Both regions' innovation-related topics are framed in two sub-fields of technology and innovation management: *research and development management*, and *technological development*. Co-citation analysis showed various journals from the canonical sources (i.e., *Research Policy*) when researching innovation (e.g., *Journal of International*

*Business Studies* or *Journal of Cleaner Production*), reinforcing the statement even in business and management studies, innovation is still multidisciplinary.

The landscape presented here would help researchers, institutions, and policymaking offices understand from a quantitative perspective the development of the research on innovation in business and management, a vertical field for economic development, enriching the framework with data from open access platforms. Discussions could shed light on the comparative results in developing countries regarding national/regional policies/incentives in both net output and incidence (i.e., citations). Also, if topics researched are central or peripheral, and to what other research-related topics would be feasible to approach (i.e., topics within *research and development management* and *technological development*). Co-authorship and co-citation networks outline high social capital researchers-institutions for strategic sharing of intangible or tangible resources for further research projects, and emergent sources of significant incidence in the field. Also, the open-access dataset could be used widely for a detailed inquiry or further replications or triangulations.

Search criteria were focused on findings research about both regions. Further research could solve this restriction by gathering research conducted by researchers affiliated with both regions to precisely understand south-south collaboration. The coverage of the bibliographic databases/search engines and the Publish and Perish API restriction could have left aside peer-review literature and gathered another type of document (i.e., theses, working papers, among others) instead. Enriching a consolidated dataset with WoS and Scopus would contribute to the fidelity and quality of bibliographic data. Finally, academic-corporate collaboration would help understand the funding or intellectual participation for driving innovativeness in business and management research.

**Acknowledgments**

[Pending: add after peer review].

## References

Aguillo, I. F. (2012). Is Google Scholar useful for bibliometrics? A webometric analysis. *Scientometrics*, *91*(2), 343–351. Scopus. https://doi.org/10.1007/s11192-011-0582-8

Baregheh, A., Rowley, J., & Sambrook, S. (2009). Towards a multidisciplinary definition of innovation. *Management Decision*, *47*(8), 1323–1339. Scopus. https://doi.org/10.1108/00251740910984578

Birkinshaw, J., Hamel, G., & Mol, M. J. (2008). Management Innovation. *Academy of Management Review*, *33*(4), 825–845. https://doi.org/10.5465/amr.2008.34421969

Casper, S., & Murray, F. (2005). Careers and clusters: Analyzing the career network dynamic of biotechnology clusters. *J Eng Technol Manage JET M*, *22*(1–2), 51–74. Scopus. https://doi.org/10.1016/j.jengtecman.2004.11.009

Ceretta, G. F., Dos Reis, D. R., & Da Rocha, A. C. (2016). Innovation and business models: A bibliometric study of scientific production on Web of Science database. *Gestao e Producao*, *23*(2), 433–444. Scopus. https://doi.org/10.1590/0104-530X1461-14

Cetindamar, D., Wasti, S. N., Ansal, H., & Beyhan, B. (2009). Does technology management research diverge or converge in developing and developed countries? *Technovation*, *29*(1), 45–58. https://doi.org/10.1016/j.technovation.2008.04.002

Choi, D. G., Lee, Y.-B., Jung, M.-J., & Lee, H. (2012). National characteristics and competitiveness in MOT research: A comparative analysis of ten specialty journals, 20002009. *Technovation*, *32*(1), 9–18. Scopus. https://doi.org/10.1016/j.technovation.2011.09.004

Confraria, H., Mira Godinho, M., & Wang, L. (2017). Determinants of citation impact: A comparative analysis of the Global South versus the Global North. *Research Policy*, *46*(1), 265–279. Scopus. https://doi.org/10.1016/j.respol.2016.11.004

Cortés-Sánchez, J. D. (2019). Innovation in Latin America through the lens of bibliometrics: Crammed and fading away. *Scientometrics*, *121*(2), 869–895. https://doi.org/10.1007/s11192-019-03201-0

Cortés-Sánchez, J. D. (2020). Innovation in China and Latin America: Bibliometric insights in business, management, accounting, and decision sciences. *Unpublished*.

Cortés-Sánchez, J. D., Bohle, K., & Guix, M. (2020). *iS in the Global South: Bibliometric study on Innovation in Management and STEM (Science, Technology, Engineering and Mathematics) for Sustainability in Developing Countries*. Unpublished.

Cui, T., & Zhang, J. (2018). Bibliometric and review of the research on circular economy through the evolution of Chinese public policy. *Scientometrics*, *116*(2), 1013–1037. https://doi.org/10.1007/s11192-018-2782-y

De Carvalho, G. D. G., Cruz, J. A. W., De Carvalho, H. G., Duclós, L. C., & De Fátima Stankowitz, R. (2017). Innovativeness measures: A bibliometric review and a classification proposal. *International Journal of Innovation Science*, *9*(1), 81–101. Scopus. https://doi.org/10.1108/IJIS-10-2016-0038

de Paulo, A. F., Carvalho, L. C., Costa, M. T. G. V., Lopes, J. E. F., & Galina, S. V. R. (2017). Mapping Open Innovation: A Bibliometric Review to Compare Developed and Emerging Countries. *Global Business Review*, *18*(2), 291–307. https://doi.org/10.1177/0972150916668600

Di Guardo, M. C., & Harrigan, K. R. (2012). Mapping research on strategic alliances and innovation: A co-citation analysis. *The Journal of Technology Transfer*, *37*(6), 789–811. https://doi.org/10.1007/s10961-011-9239-2

Dimensions. (2018). *Search*. Search. https://app.dimensions.ai/discover/publication

Fagerberg, J., Fosaas, M., & Sapprasert, K. (2012). Innovation: Exploring the knowledge base. *Research Policy*, *41*(7), 1132–1153. https://doi.org/10.1016/j.respol.2012.03.008

Falagas, M. E., Pitsouni, E. I., Malietzis, G. A., & Pappas, G. (2007). Comparison of PubMed, Scopus, Web of Science, and Google Scholar: Strengths and weaknesses. *The FASEB Journal*, *22*(2), 338–342. https://doi.org/10.1096/fj.07-9492LSF

Finardi, U. (2015). Scientific collaboration between BRICS countries. *Scientometrics*, *102*(2), 1139–1166. https://doi.org/10.1007/s11192-014-1490-5

Gao, Y., Liu, Z., Song, S., & Zheng, J. (2013). Technological capacity, product position, and firm competitiveness. *Chinese Economy*, *46*(1), 55–74. Scopus. https://doi.org/10.2753/CES1097-1475460104

Glänzel, W., & Schubert, A. (2001). Double effort = Double impact? A critical view at international co-authorship in chemistry. *Scientometrics*, *50*(2), 199–214. Scopus. https://doi.org/10.1023/A:1010561321723

Glänzel, Wolfgang, & Schubert, A. (2005). Analysing Scientific Networks Through Co-Authorship. In H. F. Moed, W. Glänzel, & U. Schmoch (Eds.), *Handbook of Quantitative Science and Technology Research: The Use of Publication and Patent Statistics in Studies of S&T Systems* (pp. 257–276). Springer Netherlands. https://doi.org/10.1007/1-4020-2755-9_12

Guan, J., & He, Y. (2007). Patent-bibliometric analysis on the Chinese science—Technology linkages. *Scientometrics*, *72*(3), 403–425. Scopus. https://doi.org/10.1007/s11192-007-1741-1

Guan, J., & Ma, N. (2007). China's emerging presence in nanoscience and nanotechnology. A comparative bibliometric study of several nanoscience "giants." *Research Policy*, *36*(6), 880–886. https://doi.org/10.1016/j.respol.2007.02.004

Gusenbauer, M. (2019). Google Scholar to overshadow them all? Comparing the sizes of 12 academic search engines and bibliographic databases. *Scientometrics*, *118*(1), 177–214. https://doi.org/10.1007/s11192-018-2958-5

Harzing, A.-W. (2007). *Publish or Perish*. Harzing.Com. https://harzing.com/resources/publish-or-perish

Harzing, A.-W., & Giroud, A. (2014). The competitive advantage of nations: An application to academia. *Journal of Informetrics*, *8*(1), 29–42. Scopus. https://doi.org/10.1016/j.joi.2013.10.007

He, T. (2009). International scientific collaboration of China with the G7 countries. *Scientometrics*, *80*(3), 571–582. https://doi.org/10.1007/s11192-007-2043-y

Herzog, C., Hook, D., & Konkiel, S. (2020). Dimensions: Bringing down barriers between scientometricians and data. *Quantitative Science Studies*, *1*(1), 387–395. https://doi.org/10.1162/qss_a_00020

Hook, D. W., Porter, S. J., & Herzog, C. (2018). Dimensions: Building Context for Search and Evaluation. *Frontiers in Research Metrics and Analytics*, *3*, 23. https://doi.org/10.3389/frma.2018.00023

Huang, L., Zhang, Y., Guo, Y., Zhu, D., & Porter, A. L. (2014). Four dimensional Science and Technology planning: A new approach based on bibliometrics and technology roadmapping. *Technological Forecasting and Social Change*, *81*(1), 39–48. https://doi.org/10.1016/j.techfore.2012.09.010

Hug, S. E., & Brändle, M. P. (2017). The coverage of Microsoft Academic: Analyzing the publication output of a university. *Scientometrics*, *113*(3), 1551–1571. https://doi.org/10.1007/s11192-017-2535-3

Jenner, F. (2019, May 15). Where China is most heavily investing in Latin America. *Latin America Reports*. https://latinamericareports.com/where-china-invest-latin-america/2064/

Jia, H. (2018). China's plan to recruit talented researchers. *Nature, S8*, 1. https://doi.org/10.1038/d41586-018-00538-z

King, D. A. (2004). The scientific impact of nations. *Nature*, *430*(6997), 311–316. https://doi.org/10.1038/430311a

Koleski, K. (2011). *Backgrounder: China in Latin America*. U.S.-China Economic and Security Review Commission. https://www.uscc.gov/research/backgrounder-china-latin-america

Kostoff, R. N. (2012). China/USA nanotechnology research output comparison-2011 update. *Technological Forecasting and Social Change*, *79*(5), 986–990. Scopus. https://doi.org/10.1016/j.techfore.2012.01.007

Kousha, K., Thelwall, M., & Rezaie, S. (2011). Assessing the citation impact of books: The role of Google Books, Google Scholar, and Scopus. *Journal of the American Society for Information Science and Technology, 62*(11), 2147–2164. https://doi.org/10.1002/asi.21608

Lazzarotti, F., Dalfovo, M. S., & Hoffmann, V. E. (2011). A bibliometric study of innovation based on schumpeter. *Journal of Technology Management and Innovation, 6*(4), 121–135. https://doi.org/10.4067/S0718-27242011000400010

Li, N., Chen, K., & Kou, M. (2017). Technology foresight in China: Academic studies, governmental practices and policy applications. *Technological Forecasting and Social Change*, *119*, 246–255. Scopus. https://doi.org/10.1016/j.techfore.2016.08.010

Li, X., Zhou, Y., Xue, L., & Huang, L. (2015). Integrating bibliometrics and roadmapping methods: A case of dye-sensitized solar cell technology-based industry in China.

*Technological Forecasting and Social Change*, *97*, 205–222. https://doi.org/10.1016/j.techfore.2014.05.007

Lopes, A. P. V. B. V., & De Carvalho, M. M. (2012). The evolution of the literature on innovation in cooperative relationships: A bibliometric study for the last two decades. *Gestao e Producao*, *19*(1), 203–217. Scopus.

Manjarrez, C. C. A., Pico, J. A. C., & Díaz, P. A. (2016). Industry Interactions in Innovation Systems: A Bibliometric Study. *Latin American Business Review*, *17*(3), 207–222. https://doi.org/10.1080/10978526.2016.1209036

Martín-Martín, A., Orduna-Malea, E., Thelwall, M., & Delgado López-Cózar, E. (2018). Google Scholar, Web of Science, and Scopus: A systematic comparison of citations in 252 subject categories. *Journal of Informetrics*, *12*(4), 1160–1177. https://doi.org/10.1016/j.joi.2018.09.002

Mueller, V., Rosenbusch, N., & Bausch, A. (2013). Success Patterns of Exploratory and Exploitative Innovation: A Meta-Analysis of the Influence of Institutional Factors. *Journal of Management*, *39*(6), 1606–1636. https://doi.org/10.1177/0149206313484516

Ng, T., & Cai, J. (2019). *China's funding for science and research to reach 2.5 per cent of GDP in 2019*. South China Morning Post. https://www.scmp.com/news/china/science/article/2189427/chinas-funding-science-and-research-reach-25-cent-gdp-2019

Padilla-Ospina, A. M., Medina-Vásquez, J. E., & Rivera-Godoy, J. A. (2018). Financing innovation: A bibliometric analysis of the field. *Journal of Business and Finance Librarianship*, *23*(1), 63–102. Scopus. https://doi.org/10.1080/08963568.2018.1448678

Pfeffer, J., & Sutton, R. (2006). *Hard facts, dangerous half-truths & total nonsense: Profiting from evidence-based management*. Harvard Business School Publishing.

Pilkington, A., & Teichert, T. (2006). Management of technology: Themes, concepts and relationships. *Technovation*, *26*(3), 288–299. https://doi.org/10.1016/j.technovation.2005.01.009

Prathap, G. (2017). Scientific wealth and inequality within nations. *Scientometrics*, *113*(2), 923–928. https://doi.org/10.1007/s11192-017-2511-y

Quan Wei, Chen Bikun, & Shu Fei. (2017). Publish or impoverish: An investigation of the monetary reward system of science in China (1999-2016). *Aslib Journal of Information Management*, *69*(5), 486–502. https://doi.org/10.1108/AJIM-01-2017-0014

Rosenbusch, N., Brinckmann, J., & Bausch, A. (2011). Is innovation always beneficial? A meta-analysis of the relationship between innovation and performance in SMEs. *Journal of Business Venturing*, *26*(4), 441–457. https://doi.org/10.1016/j.jbusvent.2009.12.002

Ryan, M. P. (2010). Patent Incentives, Technology Markets, and Public–Private Bio-Medical Innovation Networks in Brazil. *World Development*, *38*(8), 1082–1093. https://doi.org/10.1016/j.worlddev.2009.12.013

SCImago. (2018). SCImago journal ranking. *SCImago Journal Ranking*. https://www.scimagojr.com/

Silveira, F. F., & Zilber, S. N. (2017). Is social innovation about innovation? A bibliometric study identifying the main authors, citations and co-citations over 20 years. *International Journal of Entrepreneurship and Innovation Management*, *21*(6), 459–484.

Sinclair, S., & Geoffrey, R. (2012). *About Voyant Tools | Voyant Tools Documentation*. http://digihum.mcgill.ca/voyant/about/

Small, H., Sweeney, E., & Greenlee, E. (1985). Clustering the science citation index using co-citations. II. Mapping science. *Scientometrics*, *8*(5), 321–340. https://doi.org/10.1007/BF02018057

Tanco, M., Escuder, M., Heckmann, G., Jurburg, D., & Velazquez, J. (2018). Supply chain management in Latin America: Current research and future directions. *Supply Chain Management*, *23*(5), 412–430. Scopus. https://doi.org/10.1108/SCM-07-2017-0236

Thelwall, M. (2018). Dimensions: A competitor to Scopus and the Web of Science? *Journal of Informetrics*, *12*(2), 430–435. https://doi.org/10.1016/j.joi.2018.03.006

Tijssen, R. J. W. (2009). Internationalisation of pharmaceutical R&D: How globalised are Europe's largest multinational companies? *Technology Analysis and Strategic Management*, *21*(7), 859–879. Scopus. https://doi.org/10.1080/09537320903182330

Tollefson, J. (2018). China declared world's largest producer of scientific articles. *Nature*, *553*, 390–390. https://doi.org/10.1038/d41586-018-00927-4

van Eck, N. J., & Waltman, L. (2010). Software survey: VOSviewer, a computer program for bibliometric mapping. *Scientometrics*, *84*(2), 523–538. https://doi.org/10.1007/s11192-009-0146-3

Veugelers, R. (2017). The challenge of China's rise as a science and technology powerhouse. *Policy Contributions*, 15.

Waltman, L., & Larivière, V. (2020). Special issue on bibliographic data sources. *Quantitative Science Studies*, *1*(1), 360–362. https://doi.org/10.1162/qss_e_00026

Wang, K., Shen, Z., Huang, C., Wu, C.-H., Dong, Y., & Kanakia, A. (2020). Microsoft Academic Graph: When experts are not enough. *Quantitative Science Studies*, *1*(1), 396–413. https://doi.org/10.1162/qss_a_00021

Wu, L., Wang, D., & Evans, J. A. (2019). Large teams develop and small teams disrupt science and technology. *Nature*, *566*(7744), 378–382. https://doi.org/10.1038/s41586-019-0941-9

Wuchty, S., Jones, B. F., & Uzzi, B. (2007). The increasing dominance of teams in production of knowledge. *Science*, *316*(5827), 1036–1039. Scopus. https://doi.org/10.1126/science.1136099

Zhao, K. (2015). Product competition and R&D investment under spillovers within full or partial collusion games. *Latin American Economic Review*, *24*(1). Scopus. https://doi.org/10.1007/s40503-015-0018-6

Zhao, S., & Yang, H. (2008). Management practices in high-tech environments and enterprises in the People's Republic of China. *Chinese Economy*, *41*(3), 17–33. Scopus. https://doi.org/10.2753/CES1097-1475410302

Zikmund, W. G., Babin, B. J., Carr, J. C., & Griffin, M. (2010). *Business Research Methods*. South-Western Cengage Learning.

**Table 1 Query items for each bibliographic databases/search engines**

| Bibliographic databases/search engines | Query items | | | | | | |
|---|---|---|---|---|---|---|---|
| | Keyword(s) | Subject area/Field | Subject categories | Source type | Author(s) affiliation(s) | Year | Results |
| Google Scholar | Innovation; Innovación; innovação; title/abstract words: China; and each LATAM country | | | Book, Doc, PDF, HTML, Citation, N/A | | 1996-2018 | 4,354 |
| Microsoft Academic | | Business | | Journal article, proceedings paper, others | | 1996-2018 | 650 |
| Dimensions | | Business and management | | Article | | 1996-2018 | 1,202 |

Source: the authors based on Dimensions (2018). Note: data from GS and MA was gathered using Publish and Perish (Harzing, 2007). The 20 Spanish-Portuguese speaking LATAM countries considered for the query were: Argentina; Bolivia; Brazil; Chile; Colombia; Costa Rica; Cuba; Dominican Republic; Ecuador; El Salvador; Guatemala; Honduras; Mexico; Nicaragua; Panama; Paraguay; Peru; Puerto Rico; Uruguay; and Venezuela.

**Table 2 Descriptive statistics of Google Scholar, Microsoft Academic and Dimensions samples**

| | | Google Scholar | | | |
|---|---|---|---|---|---|
| China | N | 975 | | | |
| | Type | Book | 66 | 7% | |
| | | Citations | 17 | 2% | |
| | | Doc | N/A | N/A | |
| | | HTML | 20 | 2% | |
| | | PDF | 20 | 2% | |
| | | N/A | 852 | 87% | |
| | | # Authors | Citations | Cites/Author | Citer/Year |
| | Mean | 2,64 | 376,36 | 201,5 | 39,4 |
| | Median | 2 | 86 | 40 | 7,92 |
| | Min | 1 | 7 | 3 | 0,7 |
| | Max | 8 | 10,442 | 5,608 | 2,585 |
| | SD | 1,5 | 835,69 | 468,41 | 136,83 |
| LATAM | N | 3,378 | | | |
| | Type | Book | 82 | 2% | |
| | | Citations | 53 | 2% | |
| | | Doc | 7 | 0% | |
| | | HTML | 466 | 14% | |
| | | PDF | 616 | 18% | |
| | | N/A | 2154 | 64% | |
| | | # Authors | Citations | Cites/Author | Citer/Year |
| | Mean | 2,08 | 4,74 | 2,9 | 0,75 |
| | Median | 2 | 0 | 0 | 0 |
| | Min | 1 | 0 | 0 | 0 |
| | Max | 6 | 310 | 310 | 51,67 |
| | SD | 1,09 | 14,397 | 11,48 | 1,9 |
| Total | N | 4353 | | | |
| | Type | Book | 148 | 3,4% | |
| | | Citations | 70 | 1,6% | |
| | | Doc | 7 | 0,2% | |
| | | HTML | 486 | 11,2% | |
| | | PDF | 636 | 14,6% | |
| | | N/A | 3006 | 69,1% | |
| | | # Authors | Citations | Cites/Author | Citer/Year |
| | Mean | 2,2 | 87,98 | 47,38 | 9,41 |
| | Median | 2 | 1 | 1 | 0,33 |
| | Min | 1 | 0 | 0 | 0 |
| | Max | 8 | 10,442 | 5,608 | 2,585 |
| | SD | 1,2 | 424,82 | 236,78 | 66,73 |

| Microsoft Academic | | | | | | |
|---|---|---|---|---|---|---|
| China | N | | 304 | | | |
| | Type | Article | 219 | 72% | | |
| | | Proceedings | 16 | 5% | | |
| | | N/A | 69 | 23% | | |
| | | # Authors | Citations | Cites/Author | Citer/Year | |
| | Mean | 1,68 | 5,37 | 2,73 | 0,84 | |
| | Median | 1 | 0 | 0 | 0 | |
| | Min | 1 | 0 | 0 | 0 | |
| | Max | 8 | 193 | 132 | 44,5 | |
| | SD | 1,12 | 19,65 | 11,08 | 3,47 | |
| LATAM | N | | 332 | | | |
| | Type | Article | 48 | 14% | | |
| | | Proceedings | 9 | 3% | | |
| | | N/A | 275 | 83% | | |

| | | # Authors | Citations | Cites/Author | Citer/Year |
|---|---|---|---|---|---|
| | Mean | 2,24 | 0,99 | 0,55 | 0,13 |
| | Median | 2 | 0 | 0 | 0 |
| | Min | 1 | 0 | 0 | 0 |
| | Max | 8 | 56 | 56 | 4,7 |
| | SD | 1,4 | 4,08 | 3,27 | 0,45 |
| Total | N | | 636 | | |
| | Type | Article | 267 | 42% | |
| | | Proceedings | 25 | 4% | |
| | | N/A | 344 | 54% | |
| | | # Authors | Citations | Cites/Author | Citer/Year |
| | Mean | 1,98 | 3,04 | 1,5 | 0,46 |
| | Median | 1 | 0 | 0 | 0 |
| | Min | 1 | 0 | 0 | 0 |
| | Max | 8 | 193 | 132 | 44,5 |
| | SD | 1,31 | 13,93 | 8,02 | 2,42 |

| Dimensions | | | | |
|---|---|---|---|---|
| China | N | | 500 | |
| | Citations | # Authors | RCR | FCR |
| Mean | 33,66 | 2,68 | 0,006 | 8,84 |
| Median | 20 | 3 | 0 | 6,11 |
| Min | 9 | 1 | 0 | 0 |
| Max | 375 | 7 | 1,9 | 59,7 |
| SD | 38,42 | 1,22 | 0,087 | 8,57 |
| LATAM | N | | 837 | |
| | Citations | # Authors | RCR | FCR |
| Mean | 8,94 | 2,76 | 0,009 | 2,41 |
| Median | 1 | 3 | 0 | 0,38 |
| Min | 0 | 0 | 0 | 0 |
| Max | 320 | 15 | 1,9 | 43,4 |
| SD | 22,98 | 1,7 | 0,1 | 5,3 |
| Total | N | | 1337 | |
| | Citations | # Authors | RCR | FCR |
| Mean | 18,8 | 2,73 | 0,008 | 4,81 |
| Median | 9 | 3 | 0 | 2,13 |
| Min | 0 | 0 | 0 | 0 |
| Max | 375 | 15 | 1,9 | 59,7 |
| SD | 32,01 | 1,54 | 0,095 | 7,4 |

Source: the authors based on data from Google Scholar, Microsoft Academic and Dimensions extracted by Publish and Perish. Note: RCR (Relative Citation Ratio): *citation performance of an article when compared to other articles in its area of research. A value of more than 1.0 shows a citation rate above average for this];* FCR (Field Citation Ratio): *citation performance of an article, when compared to similarly-aged articles in its subject area. A value of more than 1.0 shows citation above average for this group–* and MA –type, number of authors, citations, cites/author, and cites/year.

**Table 3 Top-ten cited papers in each bibliographic databases/search engines**

| Region | Citations | Authors | Leading author affiliation | Year | Title | Source | Source h-index |
|---|---|---|---|---|---|---|---|
| | | | | | Google Scholar | | |
| China | 10,442 | Zikmund W.G., Carr J.C., Babin B., Griffin M. | N/A | 2013 | Business research methods | Book | |
| China | 8,086 | Easterby-Smith M., Thorpe R., Jackson P.R. | Lancaster University, United Kingdom | 2012 | Management research | Book | |
| China | 7,755 | Chen W., Zheng R., Baade P.D., Zhang S.,… | Deputy Director, National Office for Cancer Prevention and Control, National Cancer Center, China | 2016 | Cancer statistics in China | CA - A Cancer Journal for Clinicians | 144 |
| China | 7,125 | Collis J., Hussey R. | Brunel Business School, United Kingdom | 2013 | Business research: A practical guide for undergraduate and postgraduate students | | Book |
| China | 7,015 | Yang W., Lu J., Weng J., Jia W., Ji L., Xiao J… | Department of Endocrinology, China–Japan Friendship Hospital, China | 2010 | Prevalence of diabetes among men and women in China | New England Journal of Medicine | 933 |
| LATAM | 310 | Cimoli M. | University of Venice | 2013 | Developing Innovation Systems: Mexico in a global context | Developing Innovation Systems | Book chapter |
| LATAM | 308 | Miguel Benavente J. | Oxford University, United Kingdom | 2006 | The role of research and innovation in promoting productivity in Chile | Economics of Innovation and New Technology | 28 |
| LATAM | 263 | Ffrench-Davis R, | N/A | 2002 | Economic reforms in Chile | From Dictatorship to Democracy | Book chapter |
| LATAM | 177 | Grueso L., Rosero C., Escobar A. | N/A | 2003 | The process of black community organizing in the Southern Pacific coast region of Colombia | Cultures Of Politics/politics Of Cultures | Book chapter |
| LATAM | 151 | Newell P | School of Development Studies, University of East Anglia, United Kingdom | 2009 | Bio-hegemony: the political economy of agricultural biotechnology in Argentina | Journal of Latin American Studies | 38 |
| | | | | | Microsoft Academic | | |
| China | 193 | Zhu Y., Wittmann X., Peng M. | School of Economics and Management, Tongji University, China | 2012 | Institution Based Barriers To Innovation In Smes In China | Asia Pacific Journal of Management | 65 |
| China | 143 | Li Y., Zhao Y., Liu Y. | Management School, Xi'an Jiaotong University, China | 2006 | The Relationship Between Hrm Technology Innovation And Performance In China | International Journal of Manpower | 49 |
| China | 132 | Li M. | School of Economics & Management, Hebei University of Science and Technology, China | 2001 | China S Leap Into The Information Age Innovation And Organization In The Computer Industry | Academy of Management Review | 242 |
| China | 92 | Lazonick W. | University of Massachusetts | 2004 | Indigenous Innovation And Economic Development Lessons From China S Leap Into The Information Age | Industry and Innovation | 53 |
| China | 89 | Zheng Zhou K., Yong Gao G., Zhao H. | Faculty of Business and Economics, University of Hong Kong, Hong Kong | 2017 | State Ownership And Firm Innovation In China An Integrated View Of Institutional And Efficiency Logics | Administrative Science Quarterly | 165 |
| LATAM | 56 | Goedhuys M. | UNU-MERIT and University of Antwerpen, The Netherlands | 2007 | The Impact Of Innovation Activities On Productivity And Firm Growth Evidence From Brazil | Libri | 22 |
| LATAM | 29 | Hartwich F., Alexaki A., Baptista R. | IFPRI | 2007 | Innovation Systems Governance In Bolivia | Innovation Systems | |
| | | | | | Lessons For Agricultural Innovation Policies | Governance In Bolivia Lessons For Agricultural Innovation Policies | |
| LATAM | 26 | Chudnovsky D., Lopez A., Rossi M., Ubfal D. | Inter-American Development Bank | 2006 | Evaluating A Program Of Public Funding Of Private Innovation Activities An Econometric Study Of Fontar In Argentina | Inter-American Development Bank - Working Papers | |
| LATAM | 12 | Dahl Andersen A. | University of Oslo | 2015 | A Functions Approach To Innovation System Building In The South The Pre Proalcool Evolution Of The Sugarcane And Biofuel Sector In Brazil | Innovation and Development | 6 |
| LATAM | 12 | Alvarez E., Bravoortega C., Navarro L. | Inter-American Development Bank | 2011 | Employment Generation Firm Size And Innovation In Chile | Cepal Review | |
| | | | Dimensions | | | | |
| China | 375 | Lin H. | Department of Shipping and Transportation Management, National Taiwan Ocean University, Taiwan | 2007 | Knowledge sharing and firm innovation capability: an empirical study | International Journal of Manpower | 49 |
| China | 281 | Liu X., White S. | National Research Center of Science and Technology for Development, Research Center for Innovation Strategy and Management, China | 2001 | Comparing innovation systems: a framework and application to China's transitional context | Research Policy | 191 |
| China | 245 | Zhang Y., Li H. | Graduate School of Business, Rice University, United States | 2010 | Innovation search of new ventures in a technology cluster: the role of ties with service intermediaries | Strategic Management Journal | 232 |
| China | 215 | Yam R., Guan J., Pun K., Tang E. | Dept. of Manufacturing Engineering, City University of Hong Kong, Hong Kong | 2004 | An audit of technological innovation capabilities in chinese firms: some empirical findings in Beijing, China | Research Policy | 191 |
| China | 211 | Zhao L., Aram J. | University of Alabama, USA | 1995 | Networking and growth of young technology-intensive ventures in China | Journal of Business Venturing | 154 |
| LATAM | 320 | Giuliani E., Pietrobelli C., Rabellotti R., | University of Sussex, United Kingdom | 2005 | Upgrading in Global Value Chains: Lessons from Latin American Clusters | World Development | 150 |
| LATAM | 187 | Steffensen W., Rogers E., Speakman K. | University of Bergen, Norway | 2000 | Spin-offs from research centers at a research university | Journal of Business Venturing | 154 |
| LATAM | 180 | Bonaglia F., Goldstein A., Mathews J. | OECD Development Centre, France | 2007 | Accelerated internationalization by emerging markets' multinationals: The case of the white goods sector | Journal of World Business | 95 |
| LATAM | 165 | Altenburg T., Meyer-Stamer J. | German Development Institute, Germany | 1999 | How to Promote Clusters: Policy Experiences from Latin America | World Development | 150 |
| LATAM | 137 | Viotti E. | Senado Federal, Brazil | 2002 | National Learning Systems A new approach on technological change in late industrializing economies and evidences from the cases of Brazil and South Korea | Technological Forecasting and Social Change | 93 |

Source: the authors based on data from Google Scholar, Microsoft Academic and Dimensions extracted by Publish and Perish

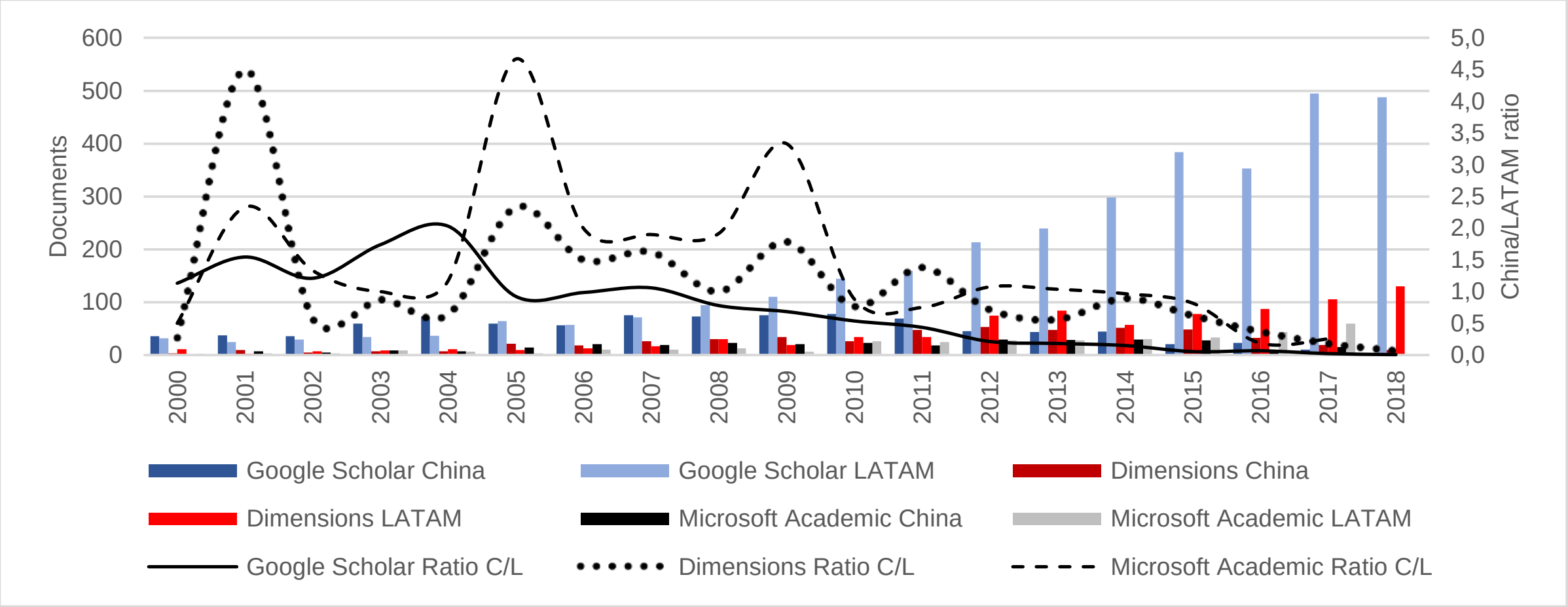

**Figure 1 China and LATAM output (left axis) and China/LATAM ratio (right axis) for Google Scholar, Dimensions, and Microsoft Academic 2001-2018. Source: the authors based on data from Google Scholar, Microsoft Academic and Dimensions extracted by Publish and Perish**

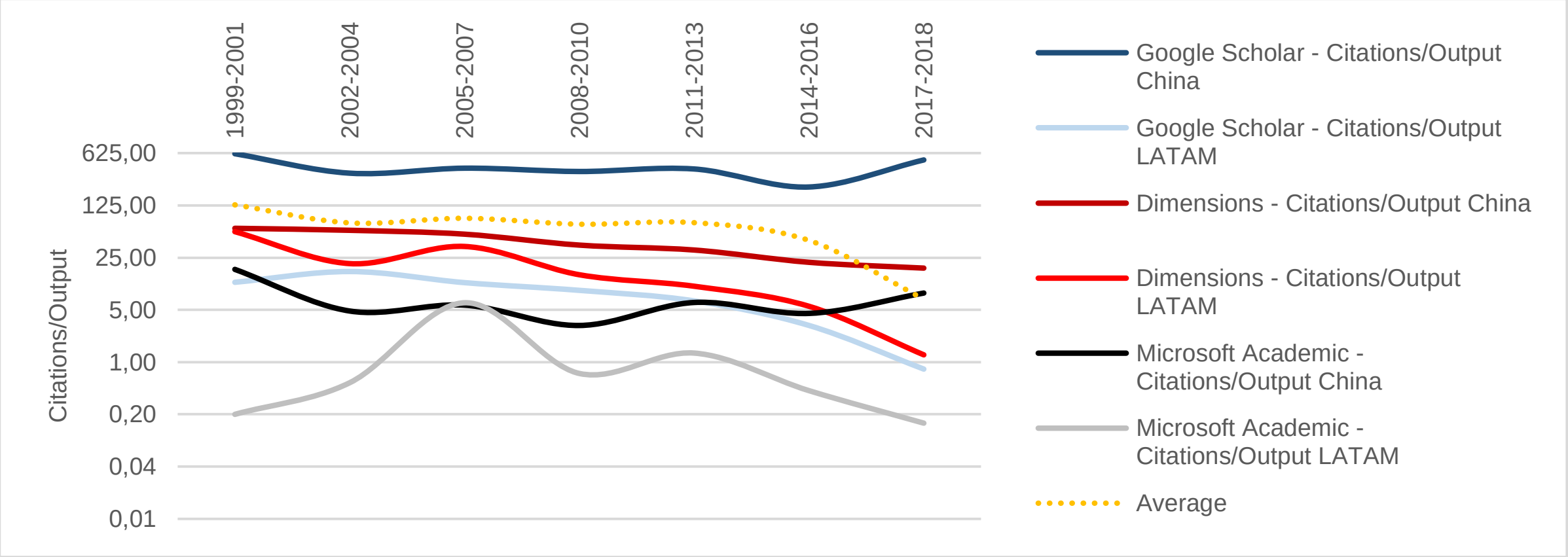

**Figure 2 China and LATAM citation/output ratio. Source: the authors based on data from Google Scholar, Microsoft Academic and Dimensions extracted by Publish and Perish**

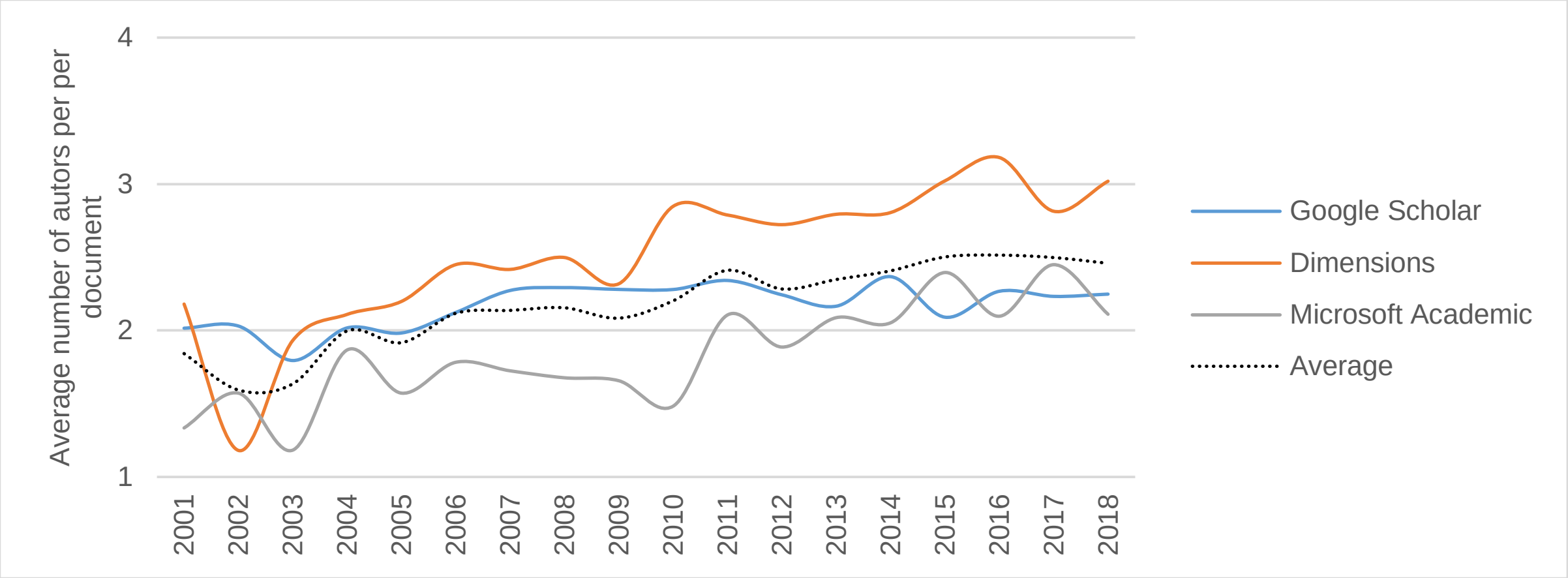

**Figure 3 Number of authors/documents 2001-2018. Source: the authors based on data from Google Scholar, Microsoft Academic and Dimensions extracted by Publish and Perish.**

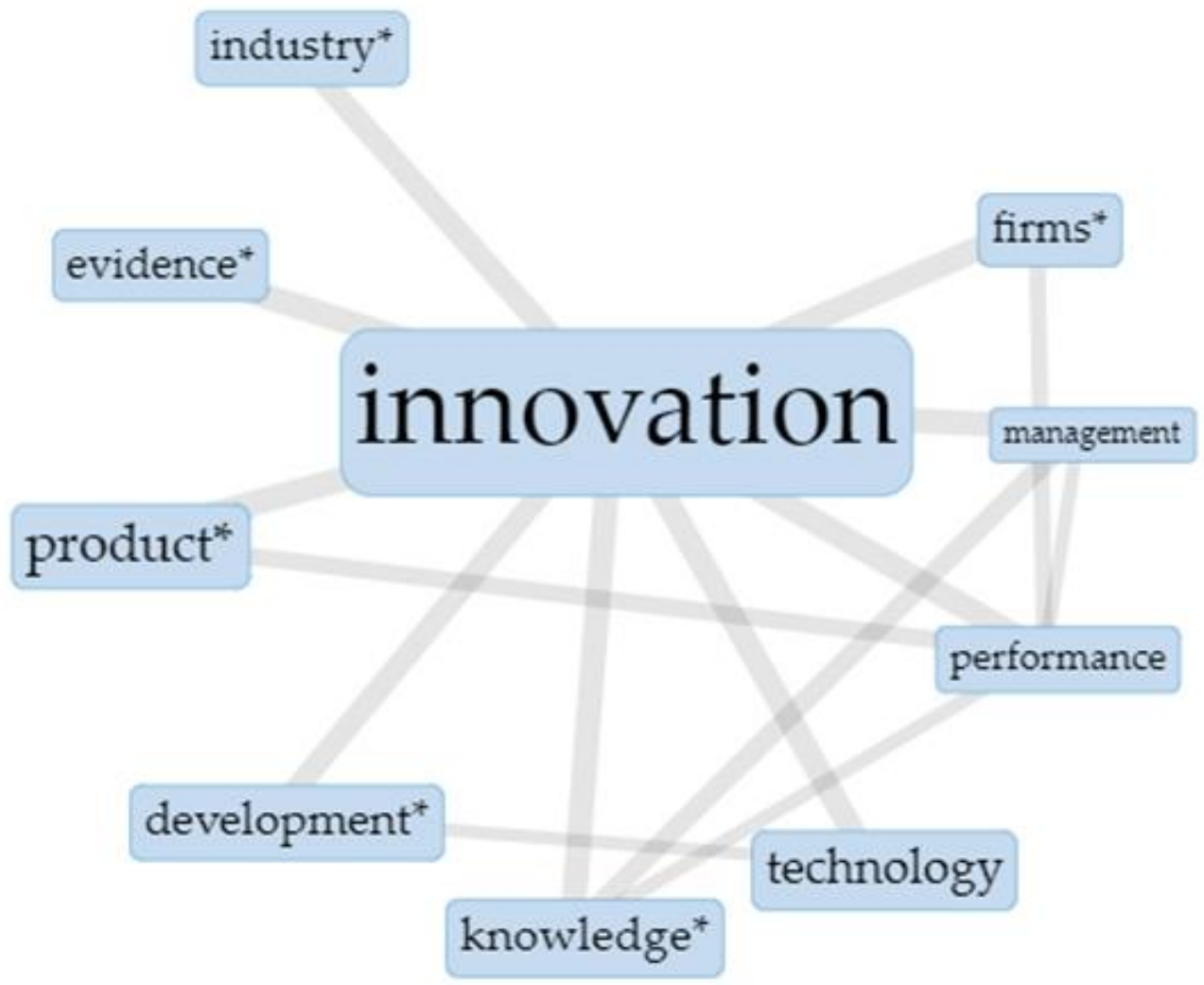

**Figure 4 Co-location graph. Source: the authors based on the complete sample of documents using Voyant-Tools.**

**China** **LATAM**

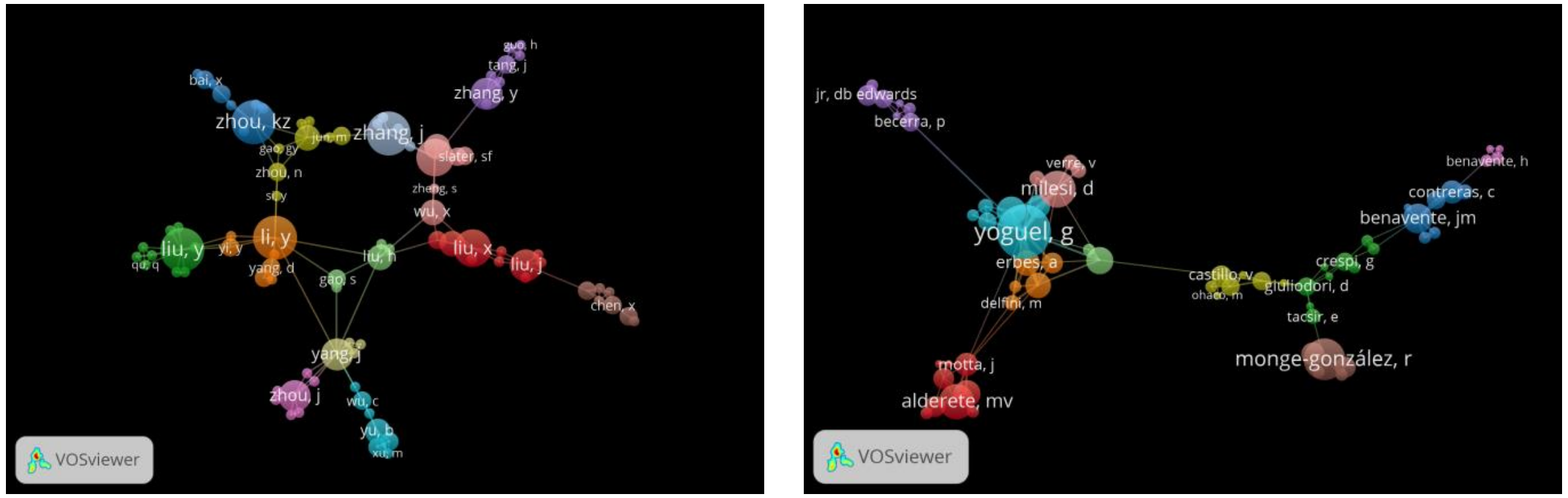

**Figure 5 Co-authorship networks of China (left) and LATAM (right). Source: the authors based on Google Scholar (2018) using VOSviewer (van Eck & Waltman, 2010).**

**China** **LATAM**

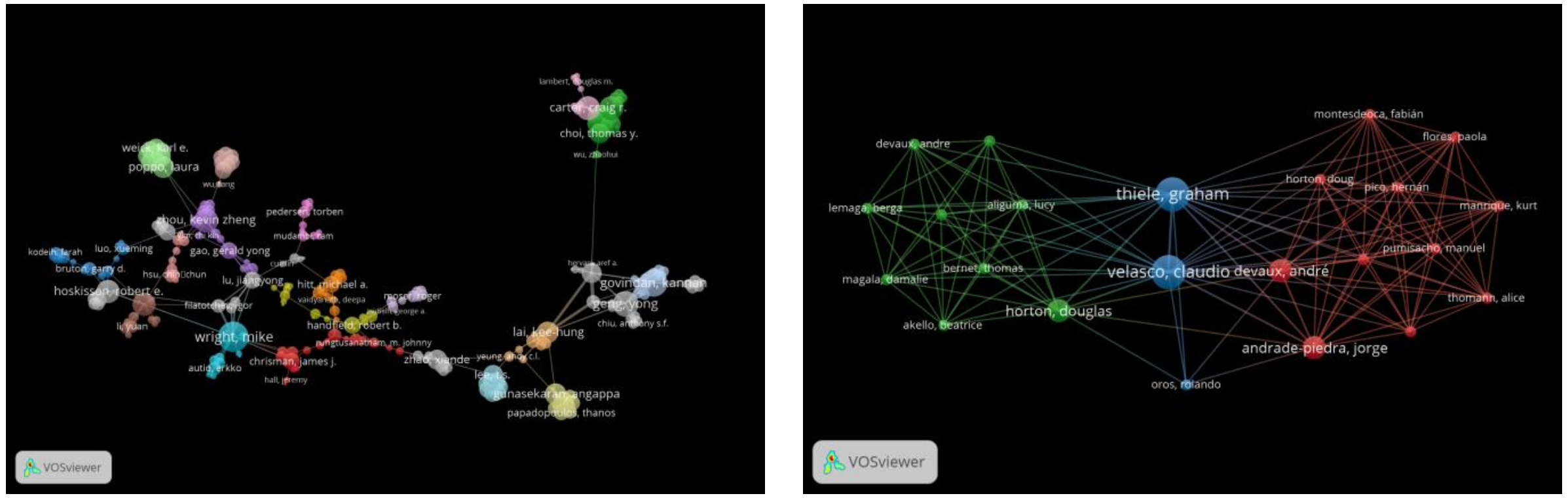

**Figure 6 Co-authorship networks of China (left) and LATAM (right). Source: the authors based on Dimensions (2018) using VOSviewer (van Eck & Waltman, 2010).**

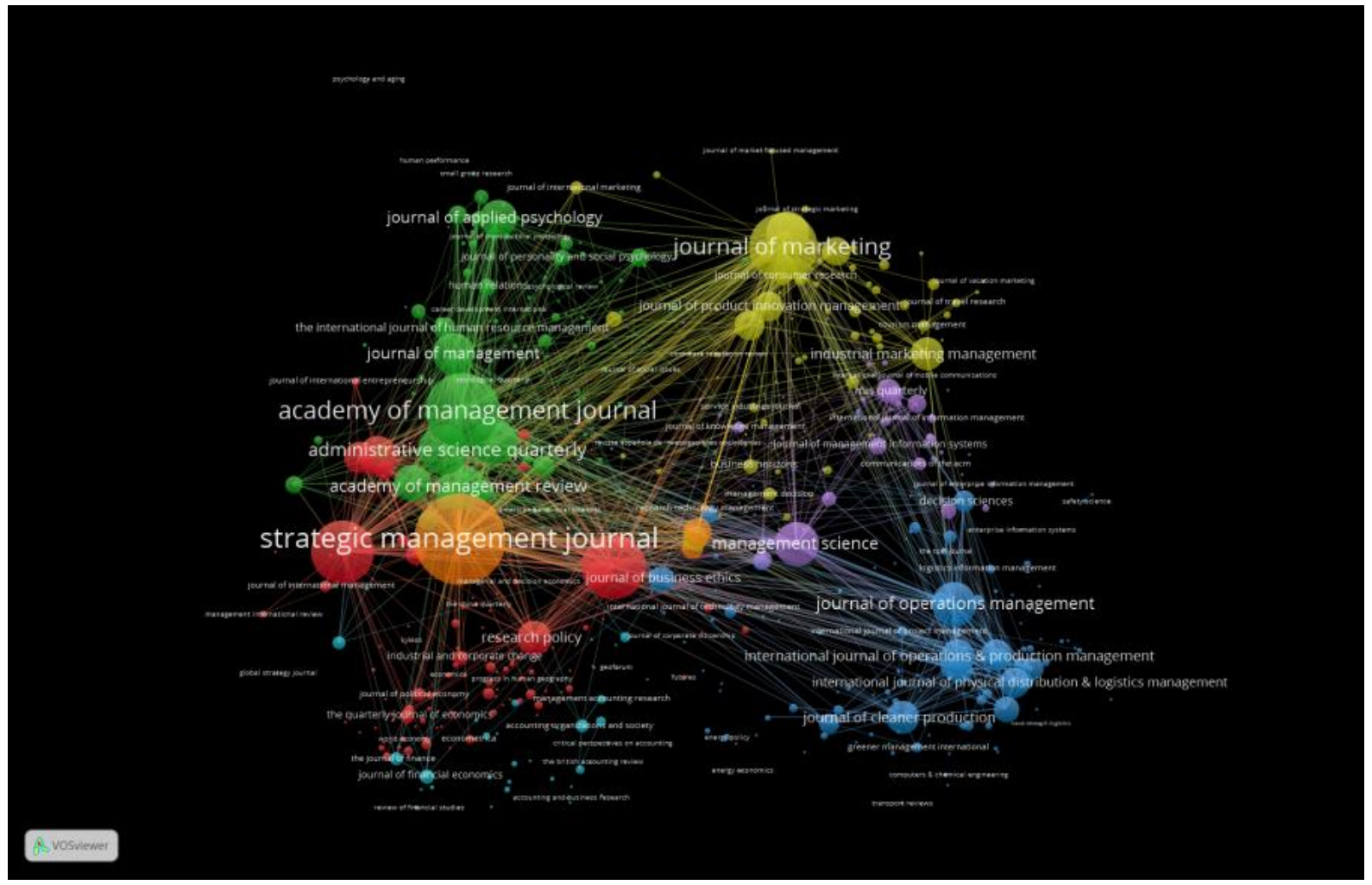

**Figure 7 Co-citation network of China. Source: the authors based on Dimensions (2018) using VOSviewer (van Eck & Waltman, 2010).**

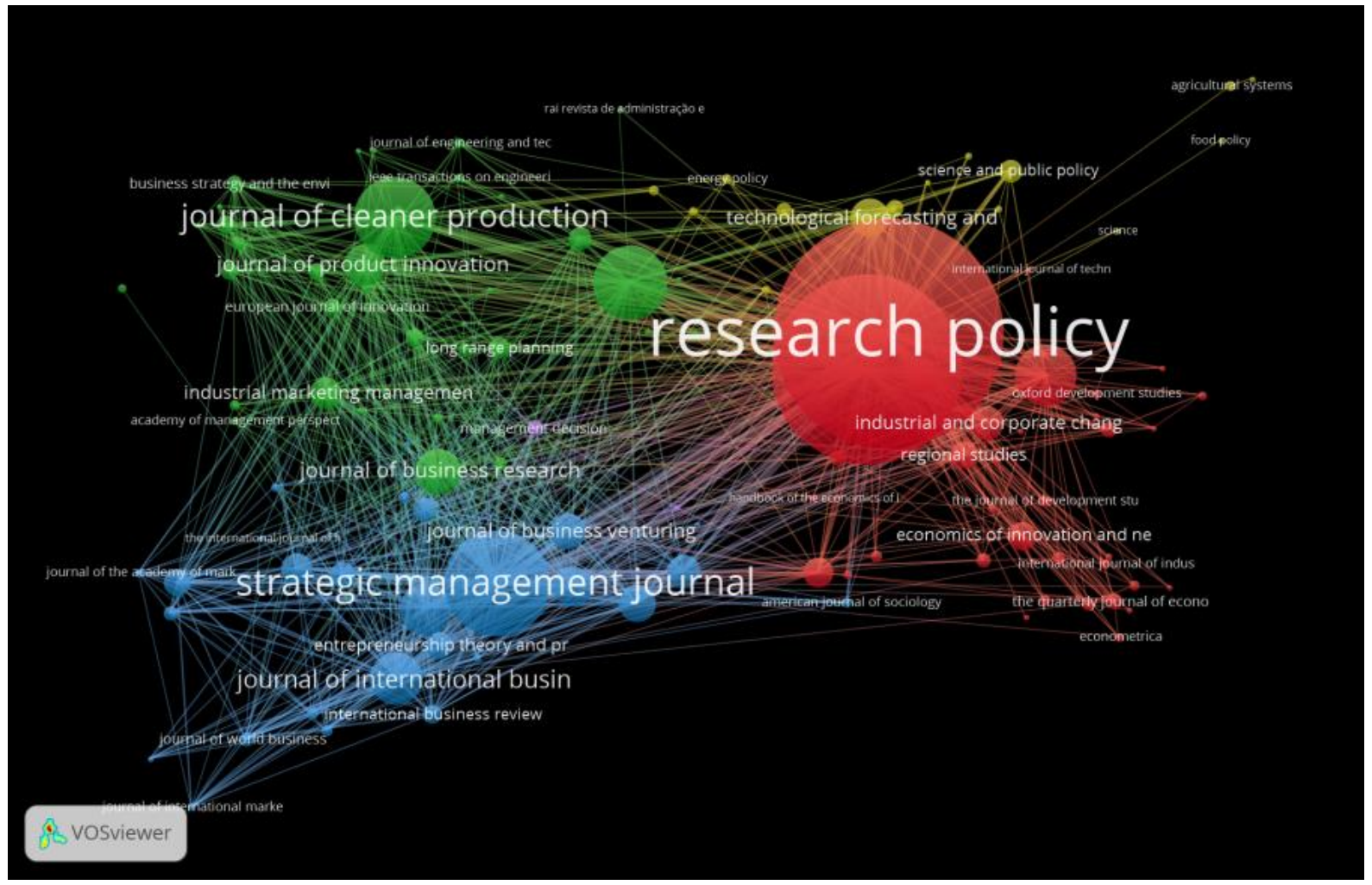

**Figure 8 Co-citation network of LATAM. Source: the authors based on Dimensions (2018) using VOSviewer (van Eck & Waltman, 2010).**